\documentclass[twocolumn,trackchanges]{aastex701}
\usepackage[utf8]{inputenc}
\usepackage{amsmath}
\usepackage{natbib}

\defcitealias{Zhang2024a}{Z24}
\defcitealias{Zou_2026}{Paper~I}

\begin{document}

\title{A systematic study of AGN feedback in a disk galaxy using MACER II: predictions of X-ray surface brightness profiles and comparison with eROSITA observations}

\correspondingauthor{Feng Yuan, Suoqing Ji}
\email{fyuan@fudan.edu.cn, sqji@fudan.edu.cn}

\author[orcid=0009-0006-4662-3053]{Yuxuan Zou}
\affiliation{Shanghai Astronomical Observatory, Chinese Academy of Sciences, 80 Nandan Road, Shanghai 200030, China}
\affiliation{University of Chinese Academy of Sciences, No. 19A Yuquan Road, Beijing 100049, China}
\email{}  

\author[orcid=0000-0003-3564-6437]{Feng Yuan} 
\affiliation{Center for Astronomy and Astrophysics and Department of Physics, Fudan University, Shanghai 200438, China}
\email{}

\author[orcid=0000-0001-9658-0588]{Suoqing Ji}
\affiliation{Center for Astronomy and Astrophysics and Department of Physics, Fudan University, Shanghai 200438, China}
\email{}

\author[orcid=0000-0002-7875-9733]{Lin He}
\affiliation{School of Astronomy and Space Science, Nanjing University, Nanjing 210023, China}
\affiliation{Key Laboratory of Modern Astronomy and Astrophysics, Nanjing University, Nanjing 210023, China}
\email{}

\author[orcid=0000-0003-0355-6437]{Zhiyuan Li}
\affiliation{School of Astronomy and Space Science, Nanjing University, Nanjing 210023, China}
\affiliation{Key Laboratory of Modern Astronomy and Astrophysics, Nanjing University, Nanjing 210023, China}
\affiliation{Institute of Science and Technology for Deep Space Exploration, Suzhou Campus, Nanjing University, Suzhou 215163, China}
\email{}

\author[orcid=0000-0001-8632-904X]{Yi Zhang}
\affiliation{Max-Planck-Institut f\"ur extraterrestrische Physik, Gie{\ss}enbachstra{\ss}e 1, 85748 Garching bei M\"unchen, Germany}
\email{}

\author[orcid=0000-0001-9200-1497]{Johan Comparat}
\affiliation{Max-Planck-Institut f\"ur extraterrestrische Physik, Gie{\ss}enbachstra{\ss}e 1, 85748 Garching bei M\"unchen, Germany}
\email{}

\author[orcid=0000-0002-2941-646X]{Zhijie Qu}
\affiliation{Department of Astronomy, Tsinghua University, Beijing 100084, PR China}
\email{}

\author[orcid=0000-0002-2853-3808]{Taotao Fang}
\affiliation{Department of Astronomy, Xiamen University, 422 Siming South Road, Xiamen, Fujian, People’s Republic of China}
\email{}
 
\begin{abstract}
Recently, we have performed a systematic study of AGN feedback in a disk galaxy within the {\it MACER} framework. Various model predictions, including the AGN duty cycle, the correlation between black hole accretion rates and star formation rates, and the (cold) gas fraction, have been compared with observations and will be presented in a series of papers. As the second paper in this series, without adjusting any model parameters, we directly use the simulation data introduced in \citetalias{Zou_2026} to compute the predicted X-ray surface brightness profile and compare it with eROSITA observations of circumgalactic medium (CGM) emission around galaxies, which provide important constraints on AGN feedback models. For this comparison, we adopt two stacked eROSITA radial profiles of X-ray surface brightness: (1) distant galaxies with $\log(M_*/M_\odot) = 10.5-11.0$ at $z\approx 0.02-0.10$ from \citet{Zhang2024a}, and (2) nearby $L^*$ galaxies within 50 Mpc from \citet{he2026}. We find that the average simulated profile over time is in good agreement with the stacked measurements of \citet{Zhang2024a} over a broad radial range (out to \(\sim 100\,\mathrm{kpc}\)). Our model predictions also match the results of \citet{he2026} at projected radii from \(\sim 20\,\mathrm{kpc}\) to $120\,\mathrm{kpc}$. While our simulations, which predict only thermal emission, are consistent with these recent X-ray observations, the limitations in our current model mean that this agreement does not preclude a potential contribution from non-thermal emission, e.g., from an extended halo of cosmic rays.
\end{abstract}


\section{Introduction}
\label{sec:intro}
The hot ($T \gtrsim 10^6$ K) circumgalactic medium (CGM) represents a dominant baryon reservoir and a critical component in galaxy evolution, tracing accretion shocks, galactic outflows, and feedback processes \citep{Spitzer1956, WhiteRees1978, WhiteFrenk1991}. For Milky Way (MW)-mass galaxies ($M_{\rm vir} \sim 10^{12} \, \mathrm{M}_{\odot}$), the origin of this hot, X-ray-emitting halo remains debated \citep{Strickland2004, Crain2010, Dave2012, Sokolowska2018, Truong2020}. Galaxy formation models predict extended hot gas heated gravitationally via accretion shocks \citep{WhiteRees1978, WhiteFrenk1991}, yet active galactic nuclei (AGN) feedback (AGN wind, jet, radiation) and stellar feedback (supernovae, stellar winds, radiation) significantly alter their properties \citep{Strickland2004, Crain2010, Truong2020}. Distinguishing the relative contributions of gravitational heating versus feedback remains challenging, particularly at the pivotal MW-mass scale where these mechanisms may intersect \citep{WechslerTinker2018}.

X-ray observations provide crucial diagnostics. While early studies detected emission concentrated near galactic planes ($\sim 1-30$ kpc scale heights) and linked luminosity ($L_X$) to galaxy properties, such as the star formation rate (SFR) \citep{Strickland2004, Tullmann2006, LiWang2013a, Wang2016}, the launch of \textit{eROSITA} has enabled the detection of faint, extended emission out to the virial radius ($\gtrsim 100$ kpc) via stacking analyses of large galaxy samples \citep{Comparat2022, Chadayammuri2022, Zhang2024a, Popesso2025I}. These observations reveal that the hot CGM in MW-mass galaxies is significantly more extended than previously thought, which poses a challenge to the existing galaxy formation models.

Several studies have compared the predictions from numerical simulations of AGN feedback and galaxy evolution with the observed X-ray surface brightness profiles \citep{Shreeram2025,Lau_2025,Grayson2025,Zhang2025}. \citet{Shreeram2025} used IllustrisTNG-based forward modeling to interpret eROSITA stacked X-ray profiles of Milky Way-mass galaxies, separating the contributions from hot CGM gas, satellites, and unresolved point sources.
They found that the inferred CGM X-ray emission is strongly dependent on the underlying halo mass distribution, with models favoring $M_{200m} \approx 3.5 \times 10^{12} M_{\odot}$ providing the best agreement with observations of \citet{Zhang2024a} (hereafter \citetalias{Zhang2024a}).
Using the Cosmology and Astrophysics with MachinE Learning Simulations (CAMELS) suite, which consists of a large number of cosmological hydrodynamical simulations with varying cosmological parameters and baryonic feedback uniquely suited for studying feedback physics at the CGM, \citet{Lau_2025} investigates how stellar and AGN feedback affect the X-ray properties of the hot CGM. They find that stronger feedback is needed than that currently implemented in the IllustrisTNG, SIMBA, and Astrid simulations to match the observed CGM properties. However, they emphasize that adopting these enhanced feedback parameters  causes deviations in the stellar mass-halo mass relations from observational constraints below the group-mass scale. 

In this work, we calculate the predicted X-ray emission of the CGM of a MW-mass galaxy based on the numerical simulation results presented in \citet{Zou_2026}, which are based on the {\it MACER} framework \citep{2018ApJ...857..121Y}, and compare the prediction with {\it eROSITA} observations. 
The paper is organized as follows. In Section \ref{sec:observe_results}, we briefly provide an overview of the main observational results, including the results from \citetalias{Zhang2024a}, and \citet{he2026}, in which a different $L_*$ sample is used. We describe the \citet{Zou_2026} simulation  in Section \ref{sec:model}. In Section \ref{subsec:model_result}, we calculate the surface brightness profile of the CGM predicted by our Fiducial model presented in \citet{Zou_2026} and compare the results with observations.  The effects of different initial conditions of our simulation are presented and discussed in Section \ref{subsec:model_result_1}. To quantitatively understand the effect of AGN feedback on controlling the CGM surface brightness profile, we have calculated the prediction of a model without AGN feedback and the results are presented in Section \ref{subsec:model_result_2}. We summarize our results in Section \ref{sec:summary}.

\section{eROSITA observational results}
\label{sec:observe_results}

\subsection{The \citet{Zhang2024a} results}

eROSITA is a highly sensitive soft X-ray telescope optimized for the \(0.3\text{--}2.3\,\mathrm{keV}\) band, making it well suited for detecting emission from the hot circumgalactic medium\citep{Merloni_2024,Predehl2021}. \citetalias{Zhang2024a} uses X-ray data from the first four SRG/eROSITA All-Sky Surveys to measure the extended X-ray emission from the hot CGM. Their analysis is based on a stacking measurement of 85,222 central galaxies in the Milky Way-mass bin (\(\log(M_*/M_\odot)=10.5\text{--}11.0\); \(M_{*,\rm med}=5.5\times10^{10}\,M_\odot\), i.e. \(\log(M_{*,\rm med}/M_\odot)\simeq10.74\); Table~3 of \citetalias{Zhang2024a}), drawn from an approximately volume-limited sample (\(0.02<z<0.10\), \(z_{\rm med}=0.08\)) selected from the SDSS DR7 spectroscopic survey.

The central galaxy sample is constructed to achieve high completeness while reducing contamination from satellite galaxies. An important consideration in the analysis is the angular resolution of eROSITA, characterized by a point-spread function (PSF) of \(\sim30^{\prime\prime}\) \citep{Merloni_2024}. This PSF sets a practical limit on the redshift range over which the extended hot CGM can be spatially resolved and, therefore, informs both the galaxy sample selection and the stacking strategy.

After masking point sources and modeling contributions from eROSITA-undetected AGN, X-ray binaries, and residual satellites, \citetalias{Zhang2024a} detects the mean CGM X-ray surface brightness profile of Milky Way-mass galaxies out to \(R_{\rm 500c}\). The profile is well described by a $\beta$-model with a slope of \(\beta = 0.43^{+0.06}_{-0.06}\), a core radius of \(6^{+9}_{-5}\,\mathrm{kpc}\), and a central surface brightness of \(\log(S_{X,0}/{\rm erg~s^{-1}~kpc^{-2}})=36.7^{+0.4}_{-0.4}\), as shown in the purple dots in the left panel of Fig.~\ref{fig:metal}.

\subsection{The \citet{he2026} results}

\citet{he2026} performs a stacking analysis of SRG/eROSITA eRASS1 data to study soft X-ray emission from the hot CGM around nearby \(L^*\) galaxies. Their sample is drawn from the 50 Mpc Galaxy Catalog (50MGC; \citealt{Ohlson_2024}) using a B-band absolute-magnitude cut \(M_{\rm B}=-20.5\pm1.0\) and a Galactic latitude cut \(|b|>15^{\circ}\) to mitigate foreground absorption. To reduce contamination from intragroup/cluster gas, they exclude galaxies in dense environments (e.g., requiring \(\log \eta_{5,{\rm LSS}}<0\), with Virgo and Fornax members removed, where \(\eta_{5,{\rm LSS}}\) is a large-scale environment (number-density) indicator based on the local density inferred from the five nearest neighboring galaxies), yielding 474 galaxies with a median distance of 36.2 Mpc.

Using exposure-corrected stacked images, with detected point and extended sources masked, \citet{he2026} detects diffuse soft X-ray emission out to \(\gtrsim 50\) kpc, while at larger radii, the measurements are consistent with upper limits. They caution that residual contributions from unresolved sources and PSF scattering may bias the inferred fluxes higher, though such contamination becomes subdominant beyond \(\sim 10\) kpc. Stacked spectra over 10--50 kpc favor a predominantly thermal component with \(kT\sim0.2\) keV and features around 0.6--0.7 keV consistent with O\,\textsc{vii}/O\,\textsc{viii} \citep{he2026}.

For comparison with our simulations, we adopt the high-mass subsample of \citet{he2026}, defined by \(\log(M_*/M_\odot)>10.38\) (237 galaxies), as our observational reference. We compare our simulated CGM profiles with their reported measurements over 10--200 kpc. Relative to the low-mass subsample, the high-mass galaxies exhibit higher X-ray intensities at nearly all radii, with the integrated 0.5--2 keV luminosity within 10--200 kpc higher by a factor of \(\sim 2.2\) \citep{he2026}.

When compared with the results of \citetalias{Zhang2024a}, the CGM surface brightness profile reported by \citet{he2026} exhibits a steeper radial decline (see black dots in Fig.~\ref{fig:metal}). The largest difference between the two measurements appears at a radius of \(\sim120\) kpc, where the surface brightness inferred by \citet{he2026} is approximately a factor of three lower than that reported by \citetalias{Zhang2024a}. This difference is consistent with the distinct sample selections and analysis approaches adopted in the two studies, including differences in distance range, spatial resolution, and the treatment of contaminating X-ray sources.

\section{The Zou et al. (2026) simulations}
\label{sec:model}
The details of the \citet{Zou_2026} simulation are presented in the first paper of this series (hereafter \citetalias{Zou_2026}). Here, we only briefly review the key features of the simulations.

\subsection{The {\it MACER} framework} \label{subsec:model_macer}

The simulations are performed within the {\it MACER} framework, which is a model designed to simulate the evolution of an individual galaxy, with a particular focus on the role of AGN feedback. The physics of {\it MACER} is described in detail in \citet{2018ApJ...857..121Y}\footnote{ Recently, the AGN-related physics introduced in MACER has been incorporated into the cosmological simulation {\it IllustrisTNG} \citep{2026arXiv260315235Z}.}. 
The model self-consistently follows the interplay among a supermassive black hole,  a dark matter halo, a stellar bulge, a stellar disk, a rotationally supported gaseous disk, and cosmological inflow. Key physical processes, such as AGN and stellar feedback, star formation, and radiative cooling, are all incorporated. 

Compared to large-scale cosmological simulations, the {\it MACER} model achieves significantly higher spatial resolution while providing a refined treatment of black hole accretion and feedback. While a comprehensive comparison is presented in \citet{2025arXiv251102796H}, we highlight several key distinctions below. The inner boundary of the simulation is set to be smaller than the Bondi radius of the accretion flow. By combining the mass flux at the inner boundary with the theory of black hole accretion, we can reliably calculate the black hole accretion rate at the event horizon, which determines the power of the AGN and is therefore essential for modeling AGN feedback. Once the mass accretion rate is obtained, the AGN outputs—namely radiation, wind, and jet—are prescribed based on accretion theory. Specifically, accretion is classified  into cold and hot modes according to the accretion rate. The properties of wind and jet (e.g., velocity and mass flux) in the hot mode are derived directly from small-scale GRMHD simulations of black hole accretion and jet formation  \citep[e.g.,][]{Yanghai_2021}, while those in the cold mode are informed directly by observations \citep[e.g.,][]{2015MNRAS.451.4169G}. 
These outputs are then injected at the inner boundary of the simulation domain, enabling a self-consistent calculation of their momentum and energy feedback on the ISM. This approach offers substantially greater precision than the phenomenological approach commonly adopted in cosmological simulations, where a fixed fraction of AGN power is often deposited into an assumed region around the AGN.


\subsection{Setup of the numerical simulation of \citetalias{Zou_2026}}
\label{subsec:sim_setup}

The simulated system in \citetalias{Zou_2026} represents an isolated, Milky Way–mass disk galaxy embedded in a fixed background gravitational potential. The galaxy consists of four main components: a central supermassive black hole (SMBH), a dark matter halo, a stellar bulge, and co-spatial stellar and gaseous disks. 
The dark matter halo follows the equilibrium profile adopted in \citet{Springel05}, with a total mass of $1.6 \times 10^{12}\,M_{\odot}$ and a concentration parameter of 12. A central SMBH with an initial mass of $5 \times 10^{7}\,M_{\odot}$ is placed at the origin. The stellar bulge and disk have masses of $1.5 \times 10^{10}\,M_{\odot}$ and $4.7 \times 10^{10}\,M_{\odot}$, respectively, consistent with a Milky Way-like galaxy \citep{Hopkins2012}. 
In the Fiducial run, star formation is treated self-consistently and converts cold gas into newly formed stars. By the end of the simulation, the final total stellar mass of $M_{\star}^{\rm final}=1.37\times10^{11}\,M_{\odot}$.

To mimic the observational stacking of many galaxies with similar stellar masses, we approximate a galaxy ensemble by a time ensemble; i.e., we average over multiple evolutionary epochs of the same simulated galaxy. We exclude rare episodes of extremely high AGN luminosity that are inconsistent with the observational sample selection. The stellar mass scale associated with our stacked quantities is the time-averaged stellar mass over the selected epochs, $\langle M_{\star}\rangle \equiv N_{\rm sel}^{-1}\sum_{i\in{\rm sel}} M_{\star}(t_i)$, for our Fiducial model $\langle M_{\star}\rangle = 1.14\times10^{11}\,M_{\odot}$


The stellar disk follows an exponential radial profile with a scale length of $3.0~\mathrm{kpc}$ and a vertical scale height of $0.3~\mathrm{kpc}$. 
The initial cold gas disk has a mass of $0.9 \times 10^{10}\,M_{\odot}$, corresponding to a gas fraction typical of star-forming disk galaxies at low redshift. 
In addition to the disk gas, the galaxy is embedded in a diffuse hot circumgalactic medium (CGM) with a total gas mass of $2.0 \times 10^{10}\,M_{\odot}$.
The initial thermal structure of the gas is bimodal: the disk interstellar medium is set to a temperature of $10^{4}~\mathrm{K}$, while the CGM is initialized at $10^{6}~\mathrm{K}$. The gas disk is placed in rotational equilibrium within the total gravitational potential, neglecting pressure support. 

Cosmological inflow is included following the prescriptions of \citet{Dekel09}. Hot-mode accretion is implemented as a quasi-spherical inflow at large radii, with a mass inflow rate of $40\,M_{\odot}\,\mathrm{yr}^{-1}$, while cold-mode accretion is modeled via filamentary inflows at intermediate radii, contributing an additional $60\,M_{\odot}\,\mathrm{yr}^{-1}$ in total.

\begin{figure*}
\centering
\includegraphics[width=0.9\textwidth]{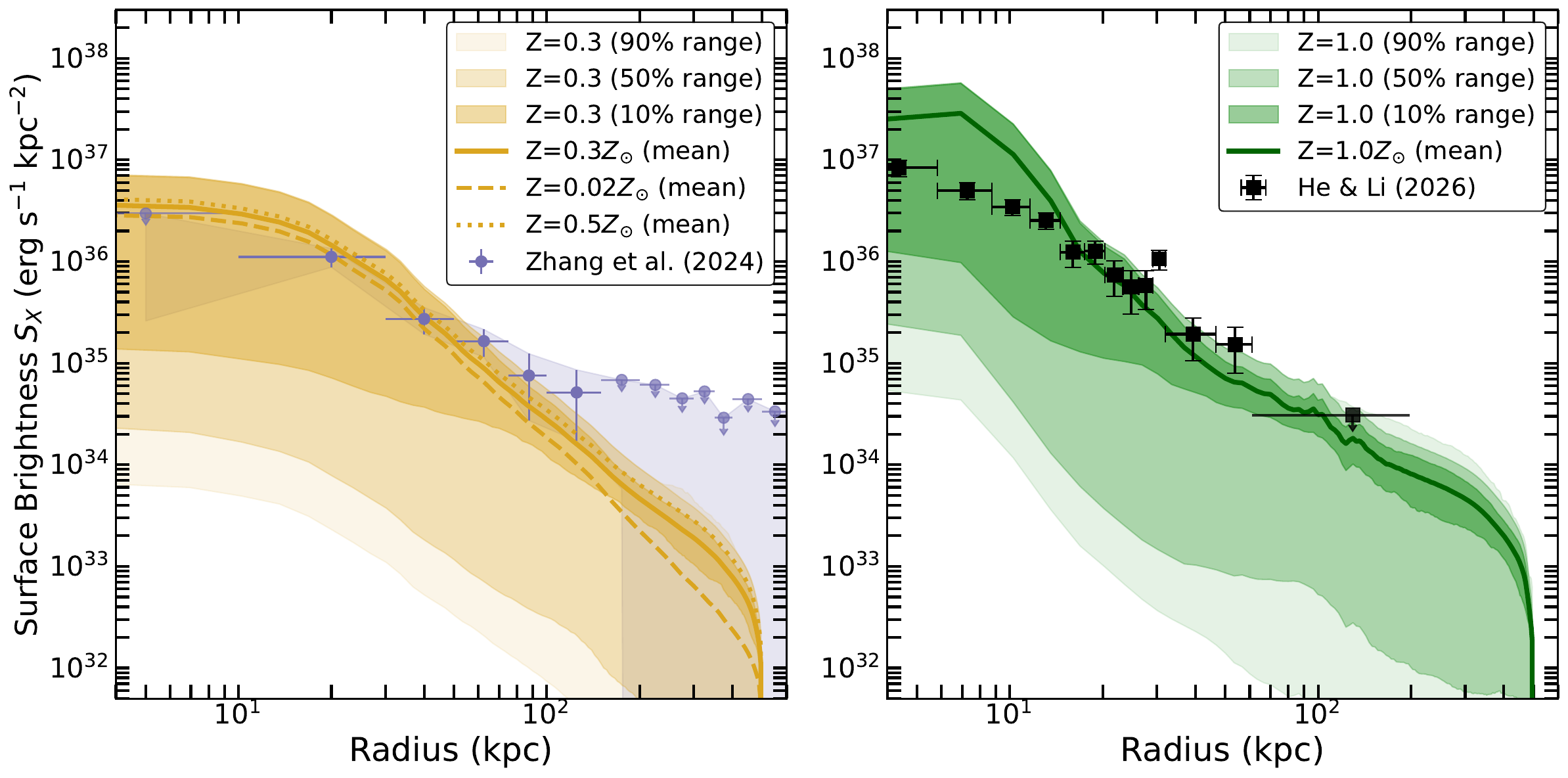}
\caption{
Simulated versus observed CGM X-ray surface-brightness profiles as a function of projected radius.
\textit{Left:} Purple points (with error bars) show the stacked profile of Milky Way-mass galaxies from \citetalias{Zhang2024a} (median distance $\sim$320~Mpc); the purple shaded region indicates the $1\sigma$ uncertainty of the best-fit $\beta$ model.
Gold curves and shaded bands show predictions of the Fiducial model from \citetalias{Zou_2026} for three gas metallicities: dashed, solid, and dotted lines correspond to $Z=0.02\,Z_\odot$, $0.3\,Z_\odot$, and $0.5\,Z_\odot$, respectively.
For $Z=0.3\,Z_\odot$, the solid curve denotes the mean profile, and the shaded regions enclose the central 10\%, 50\%, and 90\% percentile ranges.
\textit{Right:} Black squares show the stacked profile from \citet{he2026}; median distance $\sim$36~Mpc.
Because their count-rate--to--flux conversion assumes solar metallicity, we show the simulation only for $Z=1.0\,Z_\odot$.
The green solid curve is the mean Fiducial prediction, with shaded regions indicating the central 10\%, 50\%, and 90\% percentile ranges.
The simulated profile is convolved with the same angular PSF as in the left panel and rescaled to the physical scale appropriate for the closer median distance.
}
\label{fig:metal}
\end{figure*}

\begin{figure*}
\centering
\includegraphics[width=0.9\textwidth]{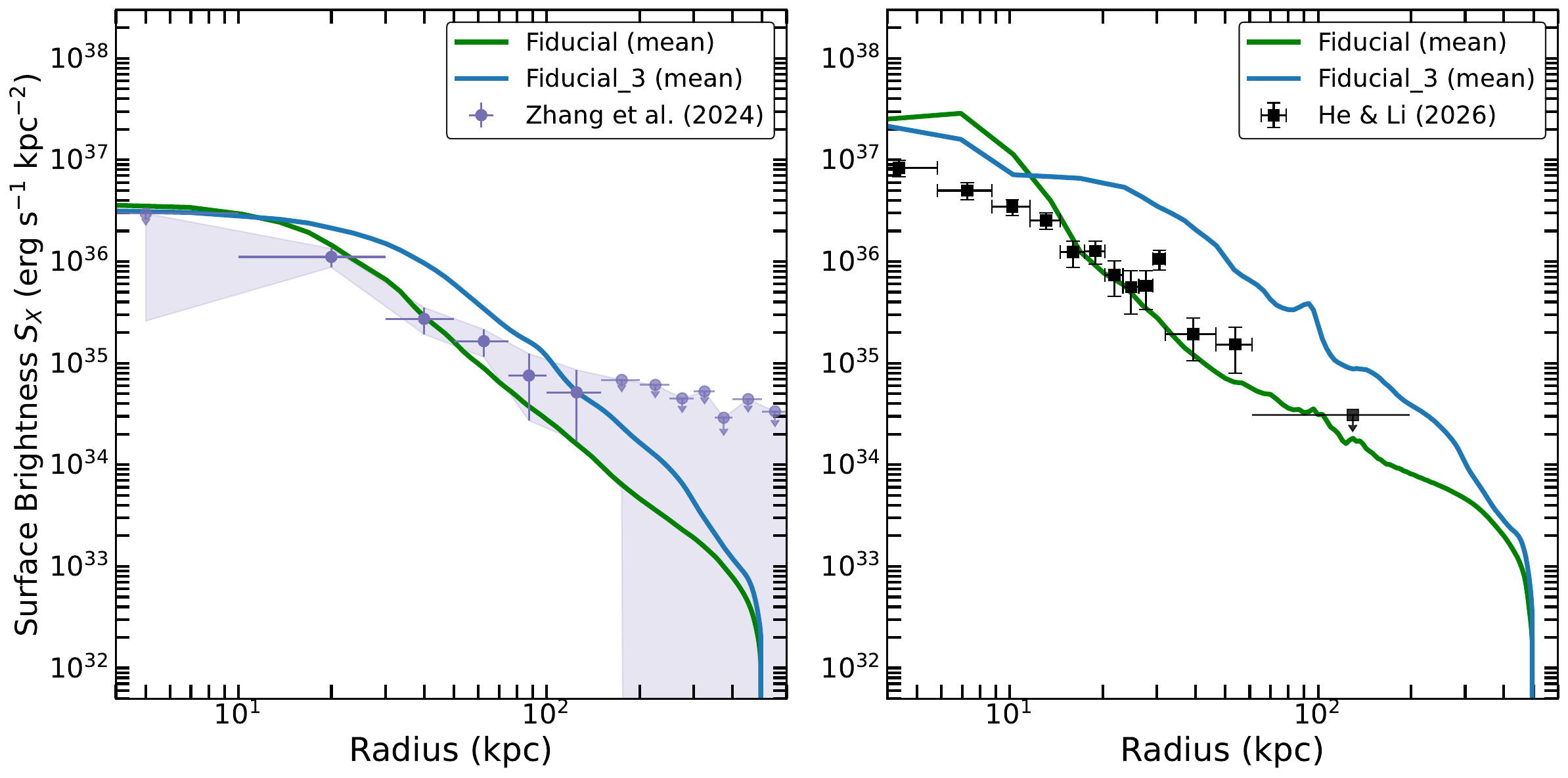}
\caption{
Comparison between simulated and observed stacked X-ray surface brightness profiles of the CGM around Milky Way-mass galaxies.
{\it Left}: Purple points with error bars show the stacked X-ray surface brightness profile from \citetalias{Zhang2024a}, whose observational sample has a median distance of $\sim$320~Mpc; the purple shaded region indicates the $1\sigma$ uncertainty of the best-fit $\beta$ model. The simulated profiles are convolved with the same angular PSF, corresponding to this median distance.
{\it Right}: Black squares show the stacked CGM X-ray surface brightness profile from He \& Li~(2025), with a median sample distance of $\sim$36~Mpc. The simulated profiles are convolved with the same angular PSF, but rescaled to the closer median distance.
In both panels, the green solid line shows the mean profile of our Fiducial model, while the blue line shows the mean profile of the Fiducial\_3 model, in which the initial CGM gas mass is three times that of the Fiducial model.
}
\label{fig:initial_c}
\end{figure*}

\begin{figure*}
\centering
\includegraphics[width=0.9\textwidth]{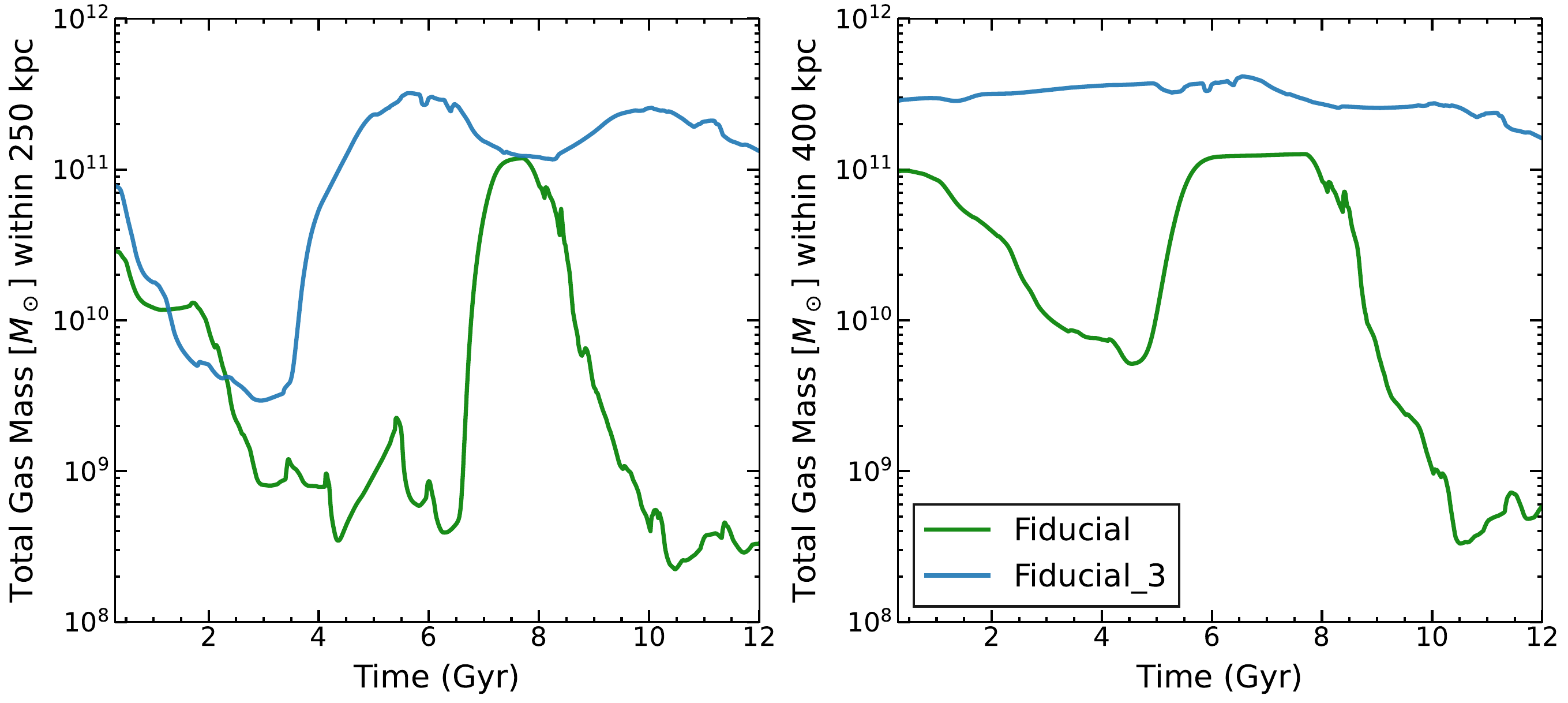}
\caption{
Time evolution of the total circumgalactic medium (CGM) gas mass in the simulations.
{\it Left}: Total gas mass enclosed within 30--250~kpc.
{\it Right}: Total gas mass enclosed within 30--400~kpc.
All gas masses are shown in units of solar masses. In both panels, the green curve corresponds to the Fiducial model, while the blue curve represents the Fiducial\_3 model, in which the initial CGM gas mass is increased by a factor of three relative to the Fiducial model.
}
\label{fig:initial_c_mass}
\end{figure*}

\section{X-ray surface brightness predictions and comparison with observations}

\subsection{The Fiducial Model}
\label{subsec:model_result}

The X-ray emissivity profile, $\epsilon_{\mathrm{X}}(r)$, is calculated using the simulation data in the 0.5--2~$\mathrm{keV}$ band as described below. For each radial bin $r$, the local X-ray volume emissivity is computed as $\epsilon_{\mathrm{X}}(r) = \Lambda(T, Z)\, n_{\mathrm{e}} n_{\mathrm{H}}$, where $n_{\mathrm{e}}$ and $n_{\mathrm{H}}$ are the electron and hydrogen number densities, $T$ is the gas temperature, $Z$ is the metallicity, and $\Lambda(T, Z)$ is the temperature- and metallicity-dependent X-ray cooling function (calculated using the APEC model). To enable direct comparison with \textit{eROSITA} observations, the intrinsic 3D emissivity profile (which is obtained by revolving the 2D simulation data around the azimuthal ($\phi$) axis under the assumption of axial symmetry) is first projected along the line of sight to generate a 2D surface brightness map $S_{\mathrm{X,intrinsic}}(x, y)$. This ideal map is then convolved with the instrumental PSF appropriate for the specific observatory and instrument configuration to produce the observable surface brightness $S_{\mathrm{X,obs}}(x, y)$.

We construct the mean radial profile in the simulations following the same pipeline adopted in the observations \citepalias{Zhang2024a}. Observationally, the mean X-ray surface brightness profile of a galaxy sample is obtained by summing the X-ray emission from individual galaxies without applying any additional weighting and then measuring a radial profile from the stacked image. To mimic this stacking procedure in our simulations, we compute the X-ray surface brightness profile for each snapshot and then average the resulting profiles. For our simulated sample, these two averaging approaches are numerically equivalent. In doing so, we exclude simulation snapshots in which the AGN luminosity is significantly higher than that characteristic of the observational comparison samples.

Since the simulations do not explicitly track metal enrichment, we assume a spatially uniform metallicity throughout the computational domain. Multi-wavelength observational constraints, including COS-Halos absorption-line measurements and XMM-Newton X-ray emission studies, indicate that the circumgalactic medium (CGM) typically exhibits sub-solar metallicities, spanning \(Z = 0.1\,Z_\odot\) to \(0.5\,Z_\odot\) with a median value of \(Z \simeq 0.3\,Z_\odot\) \citep{Werk2014,Prochaska2017,Anderson2016}. Guided by these observational results, we adopt three representative metallicities in our analysis, namely \(Z = 0.02\,Z_\odot\), \(Z = 0.3\,Z_\odot\), and \(Z = 0.5\,Z_\odot\).

When comparing our model predictions with the observational results of \citetalias{Zhang2024a}, we consider all three metallicity assumptions. This choice is motivated by the fact that their analysis is carried out at the event level, where individual X-ray photons are stacked while retaining their energy information. As a result, the X-ray surface brightness profiles are directly measured in physical units of \(\mathrm{erg\,s^{-1}\,kpc^{-2}}\), without the need to assume a spectral model for the conversion from count rate to flux. In contrast, the analysis of \citet{he2026} is performed by stacking X-ray images, which yields surface brightness profiles in terms of count rate only. In this case, converting count rate to flux necessarily requires the adoption of a spectral model; \citet{he2026} assumes solar metallicity for this conversion. For consistency, we therefore adopt solar metallicity when comparing our results with those of \citet{he2026}. Aside from the metallicity assumptions discussed above, our calculation of the X-ray surface brightness is entirely based on the simulation data of the Fiducial model presented in \citetalias{Zou_2026}.

Figure~\ref{fig:metal} presents a comparison between the simulated and observed CGM X-ray surface brightness profiles.  We exclude the interval from 7.8 to 8.2~Gyr in the \citetalias{Zou_2026} simulations to avoid epochs dominated by a luminous quasar-mode AGN phase. This exclusion is well motivated, as the AGN in both observational samples considered here---\citetalias{Zhang2024a} and \citet{he2026}---generally exhibit modest luminosities rather than sustained high-accretion states. For a fair comparison with the observations, the simulated profiles are convolved with the same angular point-spread function (PSF), with angular scales converted to physical units using the median distance of each observational sample.

We focus our quantitative comparison on projected radii beyond \(\sim 20\,\mathrm{kpc}\), where uncertainties associated with modeling unresolved eROSITA point sources in the galactic disk are minimized, as \citetalias{Zhang2024a} provides only weak constraints on the X-ray emission within \(\sim 20\,\mathrm{kpc}\). A key result is that the mean simulated profiles, averaged over metallicities \(Z = 0.02\,Z_\odot\), \(0.3\,Z_\odot\), and \(0.5\,Z_\odot\), are in good agreement with the stacked measurements of \citetalias{Zhang2024a} across a wide range of radii,  out to \(\sim 100\,\mathrm{kpc}\). In the case of \citet{he2026}, the simulations likewise reproduce the observed surface brightness profile well at radii beyond \(\sim 20\,\mathrm{kpc}\). We further note that the distributions around the mean—quantified by the central 10\%, 50\% and 90\% percentile ranges—span a relatively broad dynamic range, indicating substantial intrinsic temporal variability in the CGM X-ray emission. As a result, the time-averaged mean is weighted toward epochs of elevated surface brightness, which are typically associated with phases of enhanced AGN activity.

\subsection{Predicting the CGM mass: the role of initial conditions}
\label{subsec:model_result_1}

In observations, the total CGM mass still exhibits substantial uncertainty, mainly due to uncertainty in the mass of the hot phase, driven by the degeneracy between density, temperature, and metallicity \citep[e.g.,][]{Bregman2018, Chen2026, Zhang2026fb}.
In the Fiducial model of \citetalias{Zou_2026}, the initial mass of the CGM is set to $2\times 10^{10} M_{\odot}$. In this model, the total mass of the CGM is regulated by several processes, including mass loss due to star formation and galactic outflows and mass increase due to stellar winds and cosmological inflow. The time-averaged total mass of the CGM within 400 kpc during the galaxy's evolution, obtained from the \citetalias{Zou_2026} simulations, is $\sim 5\times 10^{10} M_{\odot}$. The CGM mass exhibits considerable temporal variation, with the interquartile range (25th to 75th percentiles) spanning from $5.2\times 10^{9}$ to $9.3\times 10^{10} M_{\odot}$. 

Different CGM masses may still give rise to similar X-ray surface brightness profiles because of variations in the underlying density and temperature distributions. To assess the sensitivity of our results to the assumed initial CGM mass and to test the robustness of our CGM mass estimate, we perform an additional simulation within the {\it MACER} framework in which the initial CGM gas mass is increased by a factor of three, to \(6\times 10^{10}\,M_\odot\) (hereafter the ``Fiducial\_3'' model). All other model parameters are held fixed; the gas metallicity is set to \(Z = 0.3\,Z_\odot\) when comparing with \citetalias{Zhang2024a}, and to \(Z = 1.0\,Z_\odot\) when comparing with \citet{he2026}.

Figure~\ref{fig:initial_c} illustrates the consequences of this modification. The time-averaged X-ray surface brightness profiles are shown in the top panels. To avoid contamination from short-lived quasar-mode activity, we exclude the interval 5.0--5.3~Gyr in the Fiducial\_3 run when computing the averages. Compared to both observational datasets, the Fiducial\_3 model predicts systematically higher X-ray surface brightness across nearly the entire CGM. In particular, relative to the stacked measurements of \citetalias{Zhang2024a}, the predicted surface brightness in the radial range \(20\text{--}100\,\mathrm{kpc}\) exceeds the observations by a factor of \(\sim 5\), while comparison with \citet{he2026} reveals an even stronger tension, with Fiducial\_3 lying well above the observed profile at essentially all radii. This systematic excess arises because the enhanced gas density in the initial conditions propagates through the subsequent galaxy evolution, leading to a persistently higher CGM density and, consequently, an elevated X-ray emissivity.

The time-averaged \emph{total} CGM gas mass predicted by the Fiducial\_3 model is 
$3\times10^{11}\,M_\odot$, approximately six times larger than that obtained in the Fiducial model of \citetalias{Zou_2026}. 
This implies that the mean total CGM mass is not simply proportional to the initial gas mass, likely due to complex feedback interplay; for example, in the higher-initial-mass case, more gas can cool and condense into cold clumps that are less susceptible to feedback, allowing a larger fraction to remain in the CGM.
Such a large total CGM mass leads to a severe overprediction of the X-ray surface brightness, thereby providing strong support for the robustness of our fiducial estimate of a typical total CGM mass of $\sim5\times10^{10}\,M_\odot$. This conclusion is consistent with recent stacked X-ray analyzes of Milky Way–mass galaxies, which indicate that CGM models with total gas masses approaching a few $\times10^{11}\,M_\odot$ would generally exceed the observed X-ray emission unless extreme assumptions about the thermal structure or metallicity are adopted \citep{zhang2025baryon}.

The right panel of Figure~\ref{fig:initial_c_mass} shows that, in the Fiducial run, the total CGM gas-mass profile within \(r \lesssim 400\,\mathrm{kpc}\) develops a pronounced high-mass plateau accompanied by a corresponding low-mass trough, with a contrast approaching two orders of magnitude. This feature originates from efficient radiative cooling, which causes a substantial fraction of the CGM gas to condense into dense cold filaments. Although this phase transition does not alter the total gas mass, it makes the gas more compact and reduces the efficiency with which AGN feedback can lift or expel it from the halo. As these cold structures subsequently migrate toward the central regions and fuel a luminous AGN phase, the ensuing strong feedback at later times eventually ejects a significant fraction of the CGM gas, giving rise to the observed mass deficit; see \citetalias{Zou_2026} for details.

Finally, in addition to our prediction for the representative CGM mass, the left panel of Figure~\ref{fig:initial_c_mass} presents the temporal evolution of the total gas mass enclosed within \(30\text{--}250\,\mathrm{kpc}\). This range is chosen to facilitate a more direct comparison with existing observational constraints: while the hot CGM can extend to larger radii, observational measurements of CGM gas mass in MW-mass galaxies are typically reported within the virial radius, which corresponds to $\sim$200–300 kpc for this mass regime. Restricting the integration to 30–250 kpc therefore places our model predictions on a more comparable footing with such constraints.

\begin{figure*}
\centering
\includegraphics[width=0.9\textwidth]{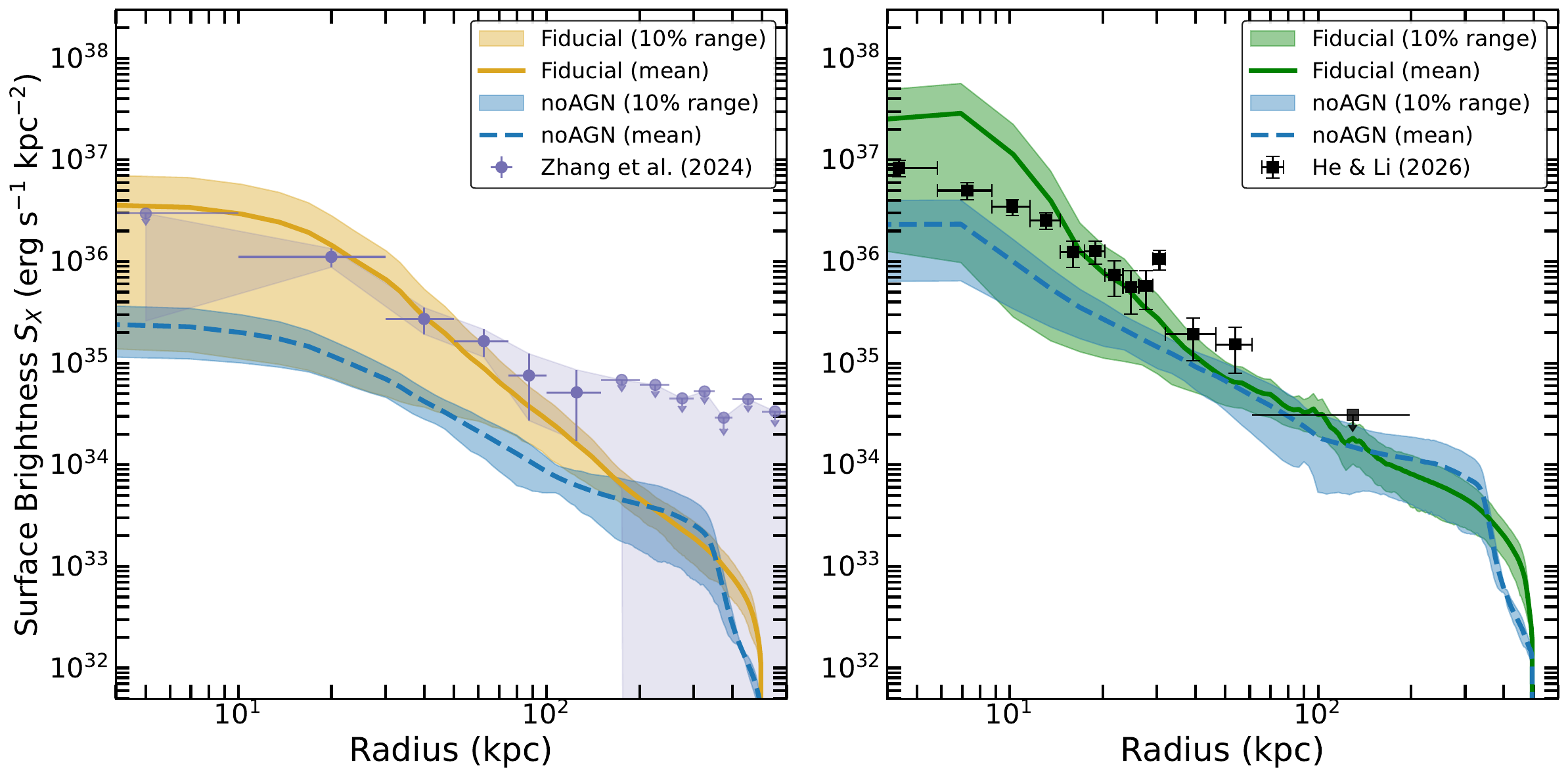}
\caption{
Comparison of simulated and observed CGM X-ray surface brightness profiles.
\textit{Left:} Purple points with error bars show the stacked X-ray surface brightness profile of Milky Way-mass galaxies from \citetalias{Zhang2024a}; the purple shaded region indicates the $1\sigma$ uncertainty of the best-fit $\beta$ model. Solid gold and dashed blue curves represent the mean predictions of the Fiducial and noAGN models, respectively; shaded bands mark the central 10\% percentile ranges.
\textit{Right:} Black squares show the stacked CGM X-ray surface brightness profile from \citet{he2026}. Solid green and dashed blue curves represent the mean predictions of the Fiducial and noAGN models, respectively, with shading again indicating the central 10\% percentile ranges.
The comparison underscores the significant role of AGN feedback in shaping the CGM X-ray emission.}
\label{fig:noagn}
\end{figure*}

\begin{figure*} 
\centering 
\includegraphics[width=\textwidth]{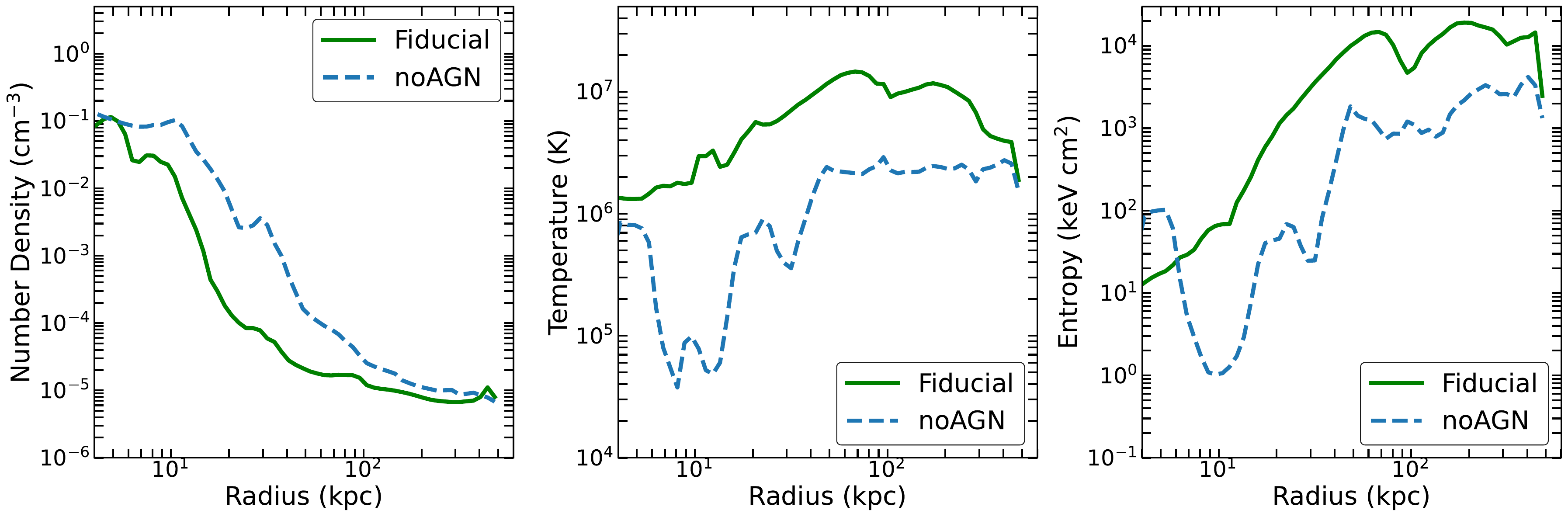} 
\caption{ 
Time-averaged radial profiles of CGM gas properties. The gas number density profile is volume weighted, while the gas temperature and gas entropy profiles are gas mass weighted.
\textit{Left:} Gas number density profile. \textit{Middle:} Gas temperature profile. \textit{Right:} Gas entropy profile. 
In all panels, the green solid curve corresponds to the Fiducial model, while the blue dashed curve represents the noAGN model. 
} 
\label{fig:radial_profile} 
\end{figure*}

\subsection{The importance of AGN Feedback} \label{subsec:model_result_2}

To investigate the role of AGN feedback in affecting the X-ray surface brightness profile, we computed the prediction of the noAGN model from \citetalias{Zou_2026}. Figure~\ref{fig:noagn} shows the comparison between the noAGN model (dashed curves) and the Fiducial model (solid curves) against two observational datasets. In the left panel, when compared with the stacked profile of Milky Way-mass galaxies from \citetalias{Zhang2024a}, the noAGN model underpredicts the X-ray surface brightness by nearly an order of magnitude from a few kiloparsecs out to $\sim 100\,\mathrm{kpc}$. In contrast, the right panel shows that, over \(R \sim 20\)--\(30\,\mathrm{kpc}\), the noAGN model falls below the Fiducial prediction only by a factor of \(\sim 2\)--3  whereas at larger radii (\(R \gtrsim 30\,\mathrm{kpc}\)) the predictions of the two models are comparable\footnote{Given the observational limitations discussed in \S\ref{subsec:model_result}, our comparison mainly focuses on the region of \(R \gtrsim 20\,\mathrm{kpc}\).}. The results clearly indicate that the physical properties of the CGM are regulated by AGN feedback: in the absence of feedback, the model predictions lie systematically below the observational data, especially in the case of the left panel of Fig.~\ref{fig:noagn}, demonstrating that AGN feedback plays a crucial role.

To understand the physical reason of the results shown in Fig. \ref{fig:noagn}, we show the time-averaged radial profiles of the number density and temperature of the CGM gas for the Fiducial run and the noAGN run in the left and middle panels of Fig. \ref{fig:radial_profile}. The AGN feedback suppresses the gas density by a factor of a few to more than an order of magnitude within \(\sim 100\,\mathrm{kpc}\), but increases the temperature by a factor of a few, from below \(10^6\,\mathrm{K}\) to \(10^6\)–\(10^7\,\mathrm{K}\) throughout the galaxy. The time-averaged radial entropy profile is correspondingly enhanced; with AGN feedback active, the entropy is higher by about one order of magnitude at essentially all radii compared to runs where AGN feedback is switched off. 

The X-ray surface brightness profile presented in Fig. \ref{fig:noagn} can be interpreted as follows. The surface brightness $S_X$
is proportional to the emission measure integrated along the line of sight:
\[
S_X \propto \int n_e^2 \,\Lambda(T,Z)\, dl,
\]
where \(\Lambda(T,Z)\) includes both bremsstrahlung continuum and metal-line emission, the latter is sensitive to metallicity. Physically, gas at \(T \sim 2 \times 10^{6}\,\mathrm{K}\) contains abundant partially ionized metals (e.g., Fe, O, and Si), which produce strong line emission through numerous electronic transitions, whereas at \(T \sim 10^{7}\,\mathrm{K}\) these elements are nearly fully stripped, so the line contribution is strongly suppressed and the free--free continuum becomes dominant. In the left panel of Fig. \ref{fig:noagn}, the Fiducial run predicts noticeably higher surface brightness than the noAGN run, even though the latter yields a higher gas density. This apparent discrepancy arises from three combined effects.  The first one is that the higher temperature in the Fiducial run enhances the bremsstrahlung emission. The second reason is that the subsolar metallicity (\(Z = 0.3\,Z_\odot\)) assumed in this panel reduces the relative contribution of line emission, making the temperature-sensitive bremsstrahlung more dominant. The third is the projection effect. The observed brightness at a given projected radius integrates emission along the entire line of sight. Significant contributions come from larger physical radii, where the densities in the two runs become similar, but the temperature (and thus the emissivity) in the Fiducial run remains substantially higher.

In the right panel, however, the difference in predicted surface brightness between the two runs is less pronounced compared to the left panel. The temperature in the noAGN run $T \sim 10^{6}\,\mathrm{K}$, so the cooling function $\Lambda(T,Z)$ enters a regime dominated by metal-line emission and becomes highly sensitive to the assumed metallicity. In the comparison with \citet{he2026} (right panel), a solar metallicity ($Z = Z_\odot$) is adopted, which substantially enhances metal-line emissivity around $T \sim 10^{6}\,\mathrm{K}$. As a result, the increase in line emission (together with the higher density) compensates for the reduced continuum at lower temperatures, so the predicted surface brightness of the noAGN model is boosted and becomes comparable to that of the Fiducial model. 


We have also examined whether a predominantly thermal CGM in our model would necessarily predict an excessively strong tSZ signal. For the Fiducial model, the time-averaged integrated tSZ signal (with the 7.8--8.2~Gyr high-AGN-luminosity phase excluded) is $Y_{200}^\mathrm{sph} \sim 2.4\times10^{-8}\ {\rm arcmin}^2$ when the galaxy is placed at a distance of 500 Mpc, with temporal scatter spanning a factor of $\sim 4$ across different epochs. This stays slightly below the current observational upper limits of tSZ stacking by \citet{das2025thermal}, while the mean value of $Y_{500}^\mathrm{sph}$ predicted by our simulations is $\sim 7.4\times10^{-8}\ {\rm arcmin}^2$, systematically lower than the observational value of $Y_{500}^\mathrm{sph} \sim 10^{-7}\,\mathrm{arcmin}^2$ reported by \citet{bregman2022extended}. During peak AGN outburst phases, $Y_{200}^\mathrm{sph}$ can reach $\sim 10^{-6}\,\mathrm{arcmin}^2$; such peak outburst values can exceed observational medians, but the duty-cycle-averaged signal remains broadly consistent within uncertainties. This result indicates that a thermal halo does not necessarily imply an overly large tSZ signal, and provides a complementary perspective to the non-thermal interpretation \citep{Ponnada2026}.

\section{Summary and discussion} \label{sec:summary}

Most recently, we have studied the evolution of a disk galaxy using {\it MACER}, a framework designed to study the effects of AGN feedback on the evolution of a single galaxy. The simulation data have been analyzed in detail, and several scientific problems have been addressed and compared with observations, including the AGN duty cycle, the correlation between black hole accretion rate and star formation rate, the galaxy quenching mechanism, and so on. The results will be presented in a series of papers.

While \citetalias{Zou_2026} in this series presents a global overview of the simulation, this second paper focuses on the surface brightness profile predicted by the Fiducial model introduced in that work. Specifically, we compute the X-ray surface brightness profile and compare it with recent eROSITA measurements of the circumgalactic medium (CGM) surrounding Milky Way–mass galaxies. All model parameters are identical to those adopted in \citetalias{Zou_2026}, and no additional free parameters are introduced. We compare our model predictions with two independent observational studies based on eROSITA data: \citetalias{Zhang2024a} and \citet{he2026}. Although both studies probe X-ray emission from the hot CGM, they differ in galaxy sample selection, redshift range, and analysis methodology. \citetalias{Zhang2024a} derives the mean CGM X-ray surface brightness profile from a large stacked sample of Milky Way–mass galaxies spanning a broad redshift range, whereas \citet{he2026} focuses on a sample of nearby ${\rm L}^*$ galaxies, providing a complementary view of the CGM on smaller physical scales. Our main results are summarized as follows.

Since there is no metallicity information in the \citetalias{Zou_2026} simulations, we assume four representative metallicities of \(Z = 0.02\,Z_\odot\), \(Z = 0.3\,Z_\odot\), \(Z = 0.5\,Z_\odot\), and \(Z = 1.0\,Z_\odot\). We find that the mean simulated surface-brightness profiles are in good agreement with the stacked measurements of \citetalias{Zhang2024a} over a wide range of radii, out to \(\sim 100\,\mathrm{kpc}\). The model predictions are also consistent with the results of \citet{he2026} at radii \(r \gtrsim 20\,\mathrm{kpc}\), extending to \(r \sim 120\,\mathrm{kpc}\) (Figure~\ref{fig:metal}). This agreement indicates that the eROSITA-detected X-ray emission could be predominantly produced by thermal radiation from hot gas, with no requirement for a significant nonthermal contribution, in line with the conclusions of \citet{he2026}. We caution, however, that this consistency does not by itself uniquely establish a thermal origin: uncertainties in the assumed initial CGM gas mass, together with the absence of cosmic rays in the current simulations, leave open the possibility that a nonthermal component contributes at some level. 

To quantitatively understand the role of AGN feedback in shaping the surface brightness, we have also calculated the surface brightness profile predicted by a model without AGN feedback (Figure \ref{fig:noagn}). We find that this model underpredicts the surface brightness by a factor of a few. The reason for the discrepancy is that the gas temperature and entropy are significantly lower than those with AGN feedback, although the gas density is higher. This indicates that the eROSITA result is a powerful tool for constraining the thermodynamic state of the gas in the galaxy and the model of AGN feedback.

We would like to point out two caveats of our work, which should be kept in mind when interpreting the comparison. First, since the simulations in \citet{Zou_2026} are performed for a single galaxy, in the present work, we approximate a galaxy ensemble with a time ensemble; that is, we stack outputs from different time steps of the same galaxy to mimic stacking different galaxies. Although real galaxies vary in dark matter halo properties, bulge-to-disk ratios, and other structural parameters, we contend that this approximation does not pose a significant problem for the present study. The X-ray surface brightness profile depends directly on gas density and temperature, the effects of other galaxy properties enter mainly through their influence on these two quantities. Given  the limited diversity in structural properties across the observed sample, we expect galaxy mass and, in particular, AGN feedback to play a more important role in determining the X-ray surface brightness profile. For example, at different epochs, AGN feedback can change the gas content of a galaxy by as much as three orders of magnitude, which would translate into an approximately six-order-of magnitude variation in the X-ray surface brightness. In this sense, our treatment should capture the zeroth-order behavior of the observational stacks.

Second, the \citet{Zou_2026} simulations span $\sim$ 12 Gyr, whereas the observations focus on low-redshift galaxies. We note that the \citet{Zou_2026} simulations are still somewhat idealized. As we emphasize in that paper, they do not account for important physical processes such as galaxy mergers and the evolution of the galactic gravitational potential driven by the cosmological growth of dark matter halos, thus the $\sim$12 Gyr evolution should not be directly compared with observations spanning different redshifts. Our goal is to isolate and investigate the key physical mechanisms that regulate galaxy evolution, rather than to obtain a precise cosmological evolution of the galaxies. Importantly, the initial gas density adopted in the \citet{Zou_2026} simulations is calibrated to match low-redshift disk galaxies. This makes the comparison with  the two observational samples consisting of galaxies at low redshift ($z \lesssim 0.1$) appropriate. 

\begin{acknowledgments}

We thank the referee and Luis C.~Ho for helpful comments and suggestions. Y.Z., F.Y., and S.J. are supported by the NSF of China (grants 12192220, 12192223, 12522301, 12133008, and 12361161601), the China Manned Space Program (grants CMS-CSST-2025-A08 and CMS-CSST-2025-A10), the National Key R\&D Program of China No. 2023YFB3002502, and the National SKA Program of China (No. 2025SKA0130100). T. F. is supported by the National SKA Program of China No. 2025SKA0150103 and the National Natural Science Foundation of China under Nos. 12550002, 12133008, 12221003, and 11890692. We acknowledge the science research grants from the China Manned Space Project with grant Nos. CMS-CSST-2021-A04 and CMS-CSST-2025-A10. Numerical calculations were run on the CFFF platform of Fudan University, the supercomputing system at the Supercomputing Center of Wuhan University, and the High Performance Computing Resource at the Core Facility for Advanced Research Computing at Shanghai Astronomical Observatory.

\end{acknowledgments}

\bibliography{sample701}{}
\bibliographystyle{aasjournalv7}

\end{document}